\documentclass{iopart}

\usepackage{graphicx,bm}
\usepackage{hyperref,amssymb}
\usepackage[numbers, sort&compress]{natbib}
\usepackage{doi}
\usepackage[textsize=tiny,backgroundcolor=yellow]{todonotes}

\hypersetup{colorlinks=true, linkcolor=blue, citecolor=blue, urlcolor=blue}
\begin{document}

\title[Self-consistent Description of the Molecule Metal-Nanoparticle Interaction]{A DFT-based Tight-Binding Approach to the Self-consistent Description of Molecule Metal-Nanoparticle Interactions}

\author{Xiaomeng Liu, Lennart Seiffert, Thomas Fennel, and Oliver K\"{u}hn$^{*}$}

\address{Institut f\"{u}r Physik, Universit\"{a}t Rostock, Albert-Einstein-Str. 23-24, D-18059 Rostock, Germany}
\ead{oliver.kuehn@uni-rostock.de} \vspace{10pt}
\begin{indented}
\item[]\today
\end{indented}

\begin{abstract}
The interaction within a hybrid system consisting of a spherical metal nanoparticle and a nearby organic dye molecule is formulated in a combined quantum-classical approach. Whereas the nanoparticle's polarization field is treated in classical multipole form, the electronic charge density of the molecule is described quantum mechanically. An efficient solution of the resulting self-consistency problem becomes possible by using the discrete representation of the charge density in terms of atom-centered Mulliken charges within the density functional theory-based tight binding (DFTB) approach. Results for two different dye molecules are presented, which focus on the dependence of the interaction on the nanoparticle's radius, the distance between nanoparticle and molecule and their mutual orientation.
\end{abstract}

%
%
\submitto{\JPB}
%
\maketitle
%
%

\section{Introduction}
Excitation of nanostructures or nanostructured surfaces with laser light enables the generation of locally confined and strongly enhanced near-fields~\cite{gramotnev14_13}. The ability to control such near-fields with nanometer spatial resolution allows to excite localized electronic processes. Popular examples that exploit this strategy include surface-enhanced Raman spectroscopy (SERS), where Raman scattering from adsorbates on a surface is extremely enhanced via near-fields generated at surface roughnesses~\cite{fleischmann74_163} or in the vicinity of metal clusters and nanoparticles~\cite{jensen07_4756,schatz07_6885,chen10_14384}, enhanced photoemission from light-harvesting complexes~\cite{wientjes14_4235}, and strong-field driven photoemission from nanostructures~\cite{zherebtsov11_656, kruger11_78, herink12_190, sussmann15_7944, seiffert16_101}. In passing we note that SERS is not limited to stationary spectroscopy, i.e.\ ultrafast nonlinear SERS  has been reported as a tool to study molecular dynamics with unprecedented sensitivity~\cite{gruenke16_2263}.  

Recent experiments on metal nanotips~\cite{kruger11_78,herink12_190} and isolated dielectric nanospheres~\cite{zherebtsov11_656,sussmann15_7944, seiffert16_101} showed that the high-energy region of measured photoelectron spectra is dominated by tunnel electrons that rescattered elastically at the nanostructures surface. In fact, the energies of these electrons can exceed the limits predicted for above threshold ionization in atomic targets~\cite{paulus94_2851} by far due to the strong acceleration in the enhanced near-fields. 
For the case of dielectric nanospheres it was shown, that the energy of emitted electrons can  even exceed  the level expected from the linear field enhancement because of the space charge fields resulting from emitted electrons and residual ions~\cite{zherebtsov11_656,sussmann15_7944, seiffert17_1096}.   
  In particular, the availability of intense laser pulses with well-defined waveforms allows to control near-field driven strong-field processes on nanometer spatial- and attosecond timescales~\cite{krausz09_163,ciappina17_054401} opening the route to realizing ultrafast light-driven nanoelectronics~\cite{krausz14_205}.

But, even in the weak field regime, field enhancement provides a means to manipulate photophysical and photochemical molecular dynamics (see, e.g., \cite{hou13_1612,piatkowski16_1401}). Of particular interest is the combination with light-driven electronic excitation energy transfer in supramolecular structures and protein-pigment complexes, which serve as antennae for light-harvesting in artificial and natural energy conversion~\cite{wurthner11_3376,schroter15_1}.  Taking a linear molecular aggregate, for instance, Frenkel excitons are usually delocalized over a range, which depends on the interaction with the fluctuating environment (nuclear degrees of freedom). Considering aggregate lengths of hundreds of nanometers, the delocalization length will usually be much smaller, i.e.\ upon absorption of a photon an exciton state on an essentially random part of the aggregate will be excited. If one is interested in the subsequent diffusive transport to some interface, e.g. in organic solar cells, the details of the initially prepared state do not really matter. However, envisioning molecular aggregates as a molecular photonic wire being part of some optical molecular device~\cite{balzani03_,tinnefeld05_217}, having a tool for  preparation of localized exciton wave packets with specific composition would be highly desirable. The local nature of the near field at a nanostructure holds the promise of providing the appropriate means for local exciton wave packet preparation. The principal feasibility has been demonstrated in a series of theoretical papers by May and coworkers~\cite{zelinskyy11_372,zelinskyy12_11330,zhang12_063533,wang17_154003}. For instance, in \cite{wang17_154003} exciton dynamics upon local excitation was systematically studied as a function of the system and laser field parameters, coming to the conclusion that excitation of local excitonic wave packets is a subtle effect, which is easily destroyed when leaving the weak field regime. For applications of local field control to photosynthetic light harvesting complexes, see, e.g., \cite{caprasecca16_2189,chen17_39720}.

A nanostructure brought into close proximity with a molecule will not only enhance the field of an incident  external laser pulse, but also react on the presence of the molecule's charge density via an additional polarization field. Vice versa, the electronic structure of the molecule will be modified in the presence of the   nanostructure. In principle the theoretical description would require to solve Maxwell's equations for the fields, self-consistently with the Schr\"odinger equation for the material system, i.e.\ nanostructure plus molecule. Since this is far from being feasible for any real system, various  approximations have been developed. May and coworkers, for instance, used a fully quantum mechanical description based on model Hamiltonians for the considered spherical metal nanoparticles (NPs) and the molecular aggregate~\cite{zelinskyy12_11330}. The plasmonic excitations of the NP are described by an effective few level system, including nonradiative decay processes. NP excitations are coupled to the aggregate via a multipole interaction. The aggregate itself is modeled by means of the Frenkel exciton Hamiltonian. The interaction with the external field is linearized, i.e. there are no field propagation effects included.

Chen et al. \cite{chen10_14384} combined classical electrodynamics of a NP, treated using the finite-difference time-domain scheme to determine a scattering response function, with a real time TDDFT description of a dye molecule to study field enhancement of absorption and Raman spectra. Thereby the backreaction of the molecular charge density  onto the NP had been neglected.  Caprasecca  et al.\ \cite{caprasecca16_2189} used a polarizable continuum model to describe a metal nanorod in terms of a complex dielectric function. This classical approach was interfaced to a linear response TDDFT calculation of molecular excitation energies via electron density dependent operators.

Real time TDDFT applied to systems as complex as molecular aggregates is computationally rather demanding. Here, we suggest to employ as an alternative the considerably more efficient DFT-based tight-binding (DFTB) method~\cite{elstner14_20120483}. During the last years, self-consistent charge DFTB has been developed to a versatile tool in Physics, Chemistry, and material science. First proposed in reference \cite{elstner98_7260}, DFTB has seen many extensions, e.g., to time-dependent problems within linear response TDDFT~\cite{niehaus01_085108} or by explicit solution of the time-dependent Kohn-Sham equations~\cite{frauenheim02_3015}. 

In this contribution, we provide a DFTB-based exploratory study of the self-consistent interaction of a NP-molecule hybrid system in the static limit. To that end we consider the simplified test case of a metal nanosphere in close proximity of a molecule. Whereas the molecular charge density is modeled via DFTB-based atom-centered Mulliken charges, the response of the nanosphere to the molecular charges is incorporated via a high-order multipole-expansion.  The paper is organized as follows: In section \ref{sec:methods}, the self-consistent description of the interaction between NP and molecule is introduced. In  section \ref{sec:results} the approach is applied to two model systems: First, the neutral tetracene ($\rm{C_{18}H_{12}}$), abbreviated as TET and, second, the positively charged 5,5',6,6'-tetraethyloro-1,1',3,3'-tetraethyl-benzimidazolyl-carbocyanine chloride ($[\rm{C_{21}H_{19}Cl_{4}N_{4}]^{+}}$), abbreviated as CCY. \textcolor{black}{The scaling behavior of the NP-molecule interaction energy with distance is analyzed in more detail in section \ref{sec:scaling}}. A summary is provided in section \ref{sec:summary}.
\section{Theoretical Methods}
\label{sec:methods}
Molecule and NP  interact with each other via the polarization field, i.e. the molecular charge density polarizes the NP whose polarization field in turn modifies the charge density.  Here, the goal is to describe this interaction in a self-consistent manner at low computational cost. Key to an efficient description is the treatment of the charge density in terms of atom-centered Mulliken charges. While the use of Mulliken charges in wavefunction or DFT-based approaches often presents a rather crude approximation to the density, they are the basic quantities of the self-consistent charge DFTB method. 

In DFTB, the molecular orbitals (MOs), $\Psi_{i}$, are expanded into the basis of atomic orbitals, $\phi_{\mu}$, according to 
\begin{equation}
  \Psi_{i}(\textbf{{r}})=\sum_{\mu}b_{\mu i}\phi_{\mu}(\textbf{{r}}-\textbf{{R}}_A) \, .
\end{equation}
Here, $\textbf{{R}}_A$ denotes the position of atom $A$ and the index $\mu= \{Alm\}$ comprises the label of the atom on which the orbital is centered, the angular momentum, and magnetic quantum number. Applying the linear variational principle, the expansion coefficients $b_{\mu i}$ are determined by the equation~\cite{elstner98_7260}
\begin{equation}
  \sum_{\nu}b_{\nu i}\left(H_{\mu\nu}-\epsilon_i S_{\mu\nu}\right)=0,\qquad \forall \mu,i \, 
\end{equation}
with $\epsilon_i$ being the energy of the $i$-th MO, $S_{\mu\nu}=\langle\phi_{\mu}\vert \phi_{\nu}\rangle$ is the overlap
integral between atomic orbitals, and the Hamiltonian matrix in the representation of the atomic orbitals is given by
\begin{eqnarray}
  \fl H_{\mu\nu}=H_{\mu\nu}^{0}+\frac{1}{2}S_{\mu\nu}\sum_C(\gamma_{AC}+\gamma_{BC})(q_C-q^0_C)+\frac{1}{2}\left(V_{A}^{\rm{ext}}+V_{B}^{\rm{ext}}\right)S_{\mu\nu}\quad \forall \mu\in A;\nu\in B \,. \nonumber\\&&
\end{eqnarray}
Here, $H_{\mu\nu}^{0}$ is the zero-order DFTB Hamiltonian matrix calculated with the atomic reference density, $q_A-q^0_A$ is the net
Mulliken charge at atom $A$ with respect to the references charge $q^0_A$, and $\gamma_{AB}$ is a measure of the
electron-electron interaction. Finally,  a local external potential
$V_{A}^{\rm{ext}}$ has been added for atom $A$. For a given external potential the MOs and their energies are determined
self-consistently, which yields respective Mulliken charges at atom $A$ according to~\cite{elstner98_7260}

\begin{equation}
  q_{A}=\frac{1}{2}\sum^{\rm{occ}}_i\sum_{\mu \in A}\sum_\nu\left(b^*_{\mu i}b_{\nu i}S_{\mu\nu}+b^*_{\nu i}b_{\mu i}S_{\nu\mu}\right)\, .
  \label{eq:mulliken}
\end{equation}
%
%
In the following we assume the coordinate origin to be at the center of the NP, which has radius $R$ and permittivity $\epsilon_r$.  Given the set of Mulliken charges $q_A$, at distances $R_A$ from the center ($A=1,\ldots,N$), the potential at distance  $r$ due to the polarization of the NP can be expressed in multipole form~\cite{seiffert18}
\begin{equation}\label{eq:polarpot}
  V_{\rm{pol}}(r)=\frac{1}{4\pi \epsilon_0}\sum_{A=1}^{N}\sum_{l=1}^{l_{\rm max}}\frac{q_A(1-\epsilon_r)l}{(1+\epsilon_r)l+1}\frac{R^{2l+1}}{r^{l+1}R_{A}^{l+1}}P_{l}(\cos\theta) \,.
\end{equation}
Here, $P_l$ is the $l$-th order Legendre polynomial, $\cos\theta$ is the directional cosine between the vectors ${\bf r}$ and ${\bf R}_A$, and the upper
summation limit $l_{\rm max}$ is chosen such as to obtain convergence with respect to changes in $V_{\rm pol}$ with a certain threshold. For the applications below this threshold is chosen as $10^{-5}$ at the smallest distance, which yields values from $l_{\rm max}=11$ for $R=3$~\AA{} to $l_{\rm max}=808$ for $R=300$~\AA. Further, the case of a metal NP is modeled using a high negative value for $\epsilon_r$ (-10$^{6}$).

\begin{figure*}[tbh]
  \begin{center}
    \includegraphics[width=0.8\textwidth]{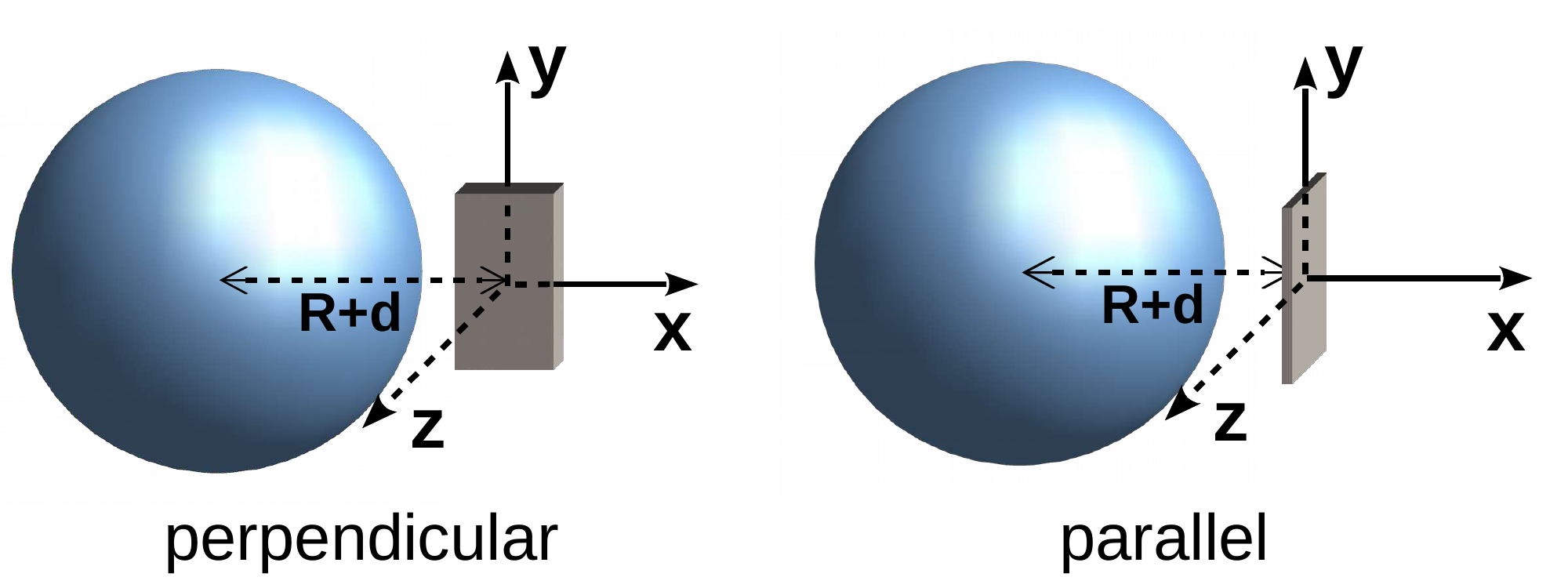}
    \caption{The perpendicular and parallel orientations of the molecule with respect to the surface of the spherical NP. $R$ is the NP's radius and $d$
      the distances from the sphere's surface to the molecule's center-of-mass. Within the respective
      planes the molecules are positioned such that the center-of-masses are at the origin of the molecule-fixed coordinate system as indicated and the principal axes point along the local $x,y,z$-axes.}
    \label{fig:orient}
  \end{center}
\end{figure*}

Since the considered molecules are essentially planar, we will discuss two geometries for the hybrid system as shown in
figure \ref{fig:orient}, i.e. the perpendicular and parallel orientation where the normal vectors of the molecule's plane and the NP surface are perpendicular and parallel to each other, respectively. The distance of the center of mass of the molecule to the surface will be called $d$.
For the description of the electronic structure of TET and CCY the dftb+ code~\cite{aradi07_5678} is used together with the halorg-0-1 Slater-Koster parameter set~\cite{elstner98_7260,kubar13_2939}. The code has been modified such as to include the potential from equation~(\ref{eq:polarpot}), which by itself is calculated with a home-made program. Given some initial set of Mulliken charges obtained without the NP, potential and Mulliken charges are iteratively adjusted until self-consistency. The threshold for the latter was set to $10^{-5}$~a.u. for changes in $V_{\rm{pol}}(r)$ summed over all atomic positions. Below we will discuss the self-consistent polarization potential  $V_{\rm pol}^{\rm SC}(r)$.

In the following, the effect of polarization of the molecule due to the NP will be investigated in dependence on the orientation and distance of the molecule and the radius of the NP. The analysis will be based on the change of Mulliken charges due to the interaction with the NP, i.e. $\Delta q_A = q_A - q_A^{\rm free}$, where $q_A^{\rm free}$ are the Mulliken charges of the bare molecule. As a global measure the root-mean-squared deviation
\begin{equation}
  \Delta q=\sqrt{\sum_{A}\left(q_{A}-q_{A}^{\rm{free}}\right)^2}
\end{equation}
will be used. Further, we will give the change in dipole moment $\Delta {\bm \mu}={\bm \mu}-{\bm \mu}^{\rm free}$ and  in total energy, $\Delta E= E-E^{\rm{free}}$, of the molecule in the polarization potential of the NP.  Here, ${\bm \mu}^{\rm free}/E^{\rm{free}}$ corresponds to the case without the NP. Concerning the energy change, i.e. the interaction energy, results from the solutions of the DFTB equations will be compared with the classical expression for the energy of the Mulliken charges in the polarization potential
\begin{equation}
\label{eq:epol}
	E_{\rm pol} = \sum_A q_A V^{\rm SC}_{\rm pol}(R_A) \, .
\end{equation}
\textcolor{black}{
Finally, we would like to emphasize that with decreasing radius the atomistic nature of the NP comes into play. If the distance between atoms at the surface of a small NP, i.e. a cluster, is comparable to the size of the molecule and their mutual distance is small, the polarization field will acquire a complicated structure and deviations from the results presented here are to be expected.
}
\section{Results}
\label{sec:results}
\subsection{Characterization of Model Systems}
\begin{figure*}[tbh]
  \begin{center}
    \includegraphics[width=0.9\textwidth]{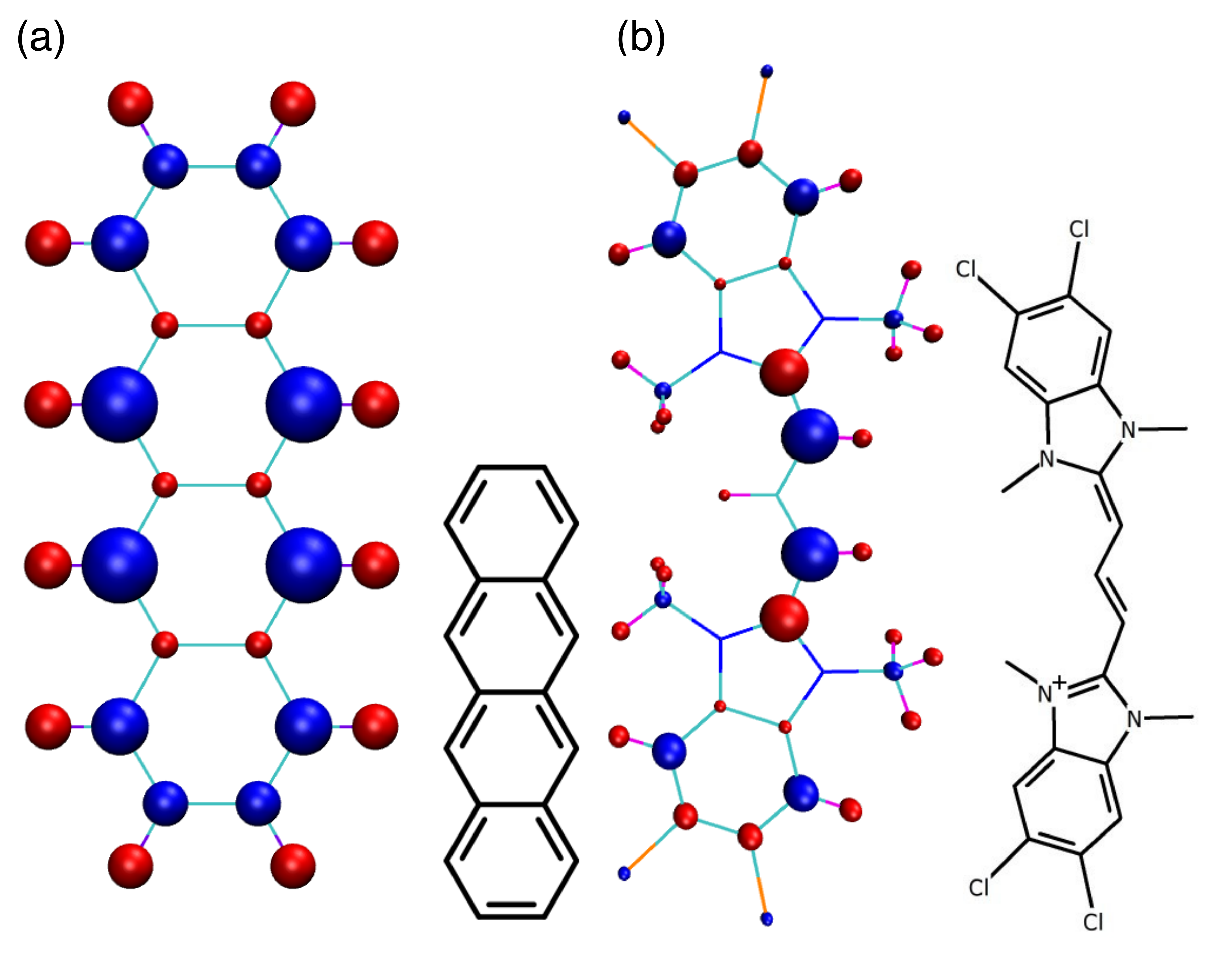}
    \caption{Mulliken charges for free neutral TET (a) and cationic CCY (b) together with the respective chemical structures. Blue and red color refer to negative and positive charge, respectively. The radii of the spheres are linearly proportional to the magnitude of the respective charge. The maximum negative charge is -0.123~$e$ and -0.273~$e$ for panel (a) and (b), respectively.}
    \label{fig:model_sys}
  \end{center}
\end{figure*} 
The effects of polarization of  a molecular charge density in the presence of a  NP will be illustrated for two model systems typically used in the context of molecular aggregates and crystals. First, the charge neutral tetracene (TET), figure \ref{fig:model_sys}a, is considered as being a representative of molecular crystal forming simple aromatic hydrocarbons~\cite{schwoerer07_}. Second, we have chosen a cationic carbocyanine (CCY) dye, i.e. a simple derivative of the 5,5',6,6'-tetrachlorobenzimidacarbocyanine (TBC) chromophore, see figure \ref{fig:model_sys}b. TBC derivatives have been widely studied~\cite{aradi07_5678,karaca11_160} and show, in particular, a rich aggregation behavior depending on the counter ion and pH, solvent and type of derivative~\cite{vonberlepsch12_119,v.berlepsch13_4948}. Note that, in order to focus on polarization effects due to a charged species, we do not consider counter ions here. Although there have been some reports of TBC-inorganic hybrid systems~\cite{halpert09_9986}, the present study does not focus on the practical realization of such systems, but on the principal effect of a nearby NP on the molecular charge density.

The Mulliken charges at the DFTB equilibrium geometry in the electronic ground state are shown in figure \ref{fig:model_sys}. In contrast to TET (panel a), the pattern of Mulliken charges in CCY (panel b) is considerably more structured. The two benzimidazolo moieties carry a net charge of 0.65~$e$ each, whereas the trimethine bridge has a net charge of -0.3~$e$.  The HOMOs of both molecules reflect the extended $\pi$ system (not shown) for the planar geometries (apart from methyl groups in case of CCY which conform to the C$_s$ symmetry.) TET has no permanent dipole moment, whereas for CCY it is oriented along the short axis with a magnitude of 1.1~D. Finally, the dimension of the $\pi$-system is about 11$\times$5~\AA{} for TET and about 17$\times$6~\AA{} for CCY; in the following the lengths of the long axes will be called $L_{\rm mol}$.
\subsection{Case of Tetracene}
\begin{figure*}[tbh]
  \begin{center}
    \includegraphics[width=0.7\textwidth]{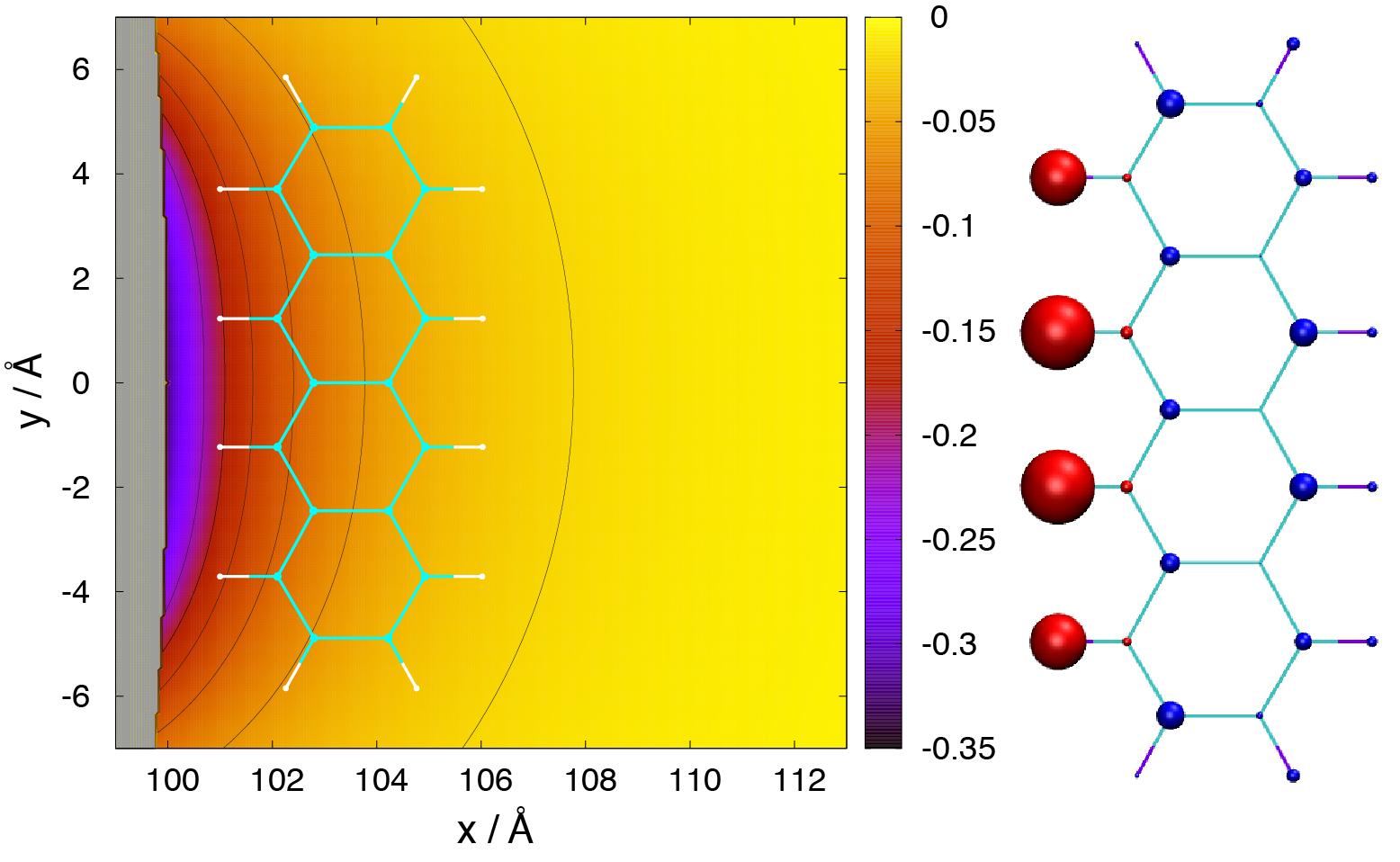}
    \caption{Left: The self-consistent polarization potential $V_{\rm pol}^{\rm SC}(r)$ (color bar in eV, contour values: -0.22, -0.18, -0.14, -0.1, -0.06, -0.02) for TET in perpendicular orientation and for $R=100\ \rm{\AA}$ and       $d=3.5\ \rm{\AA}$ (The grey area corresponds to the NP.).  Right: The change of Mulliken charges $\Delta q_A$; maximum at 0.0039~$e$.  }
    \label{fig:TET_xy_pot}
  \end{center}
\end{figure*}

 \Fref{fig:TET_xy_pot} (left) shows the polarization potential  for the perpendicular orientation  and for a NP with radius $R=100\ \rm{\AA}\gg L_{\rm mol}$  and $d=3.5\ \rm{\AA}$.   The maximum potential difference is about 0.15~eV across the short axis of the molecule. The change in the electron density as expressed by the Mulliken charges is very small (right panel). It reaches a maximum value of 0.004~$e$ or 5\% for the hydrogen atoms, which are closest to the NP.
 
\begin{figure*}[tbh]
  \begin{center}
    \includegraphics[width=0.7\textwidth]{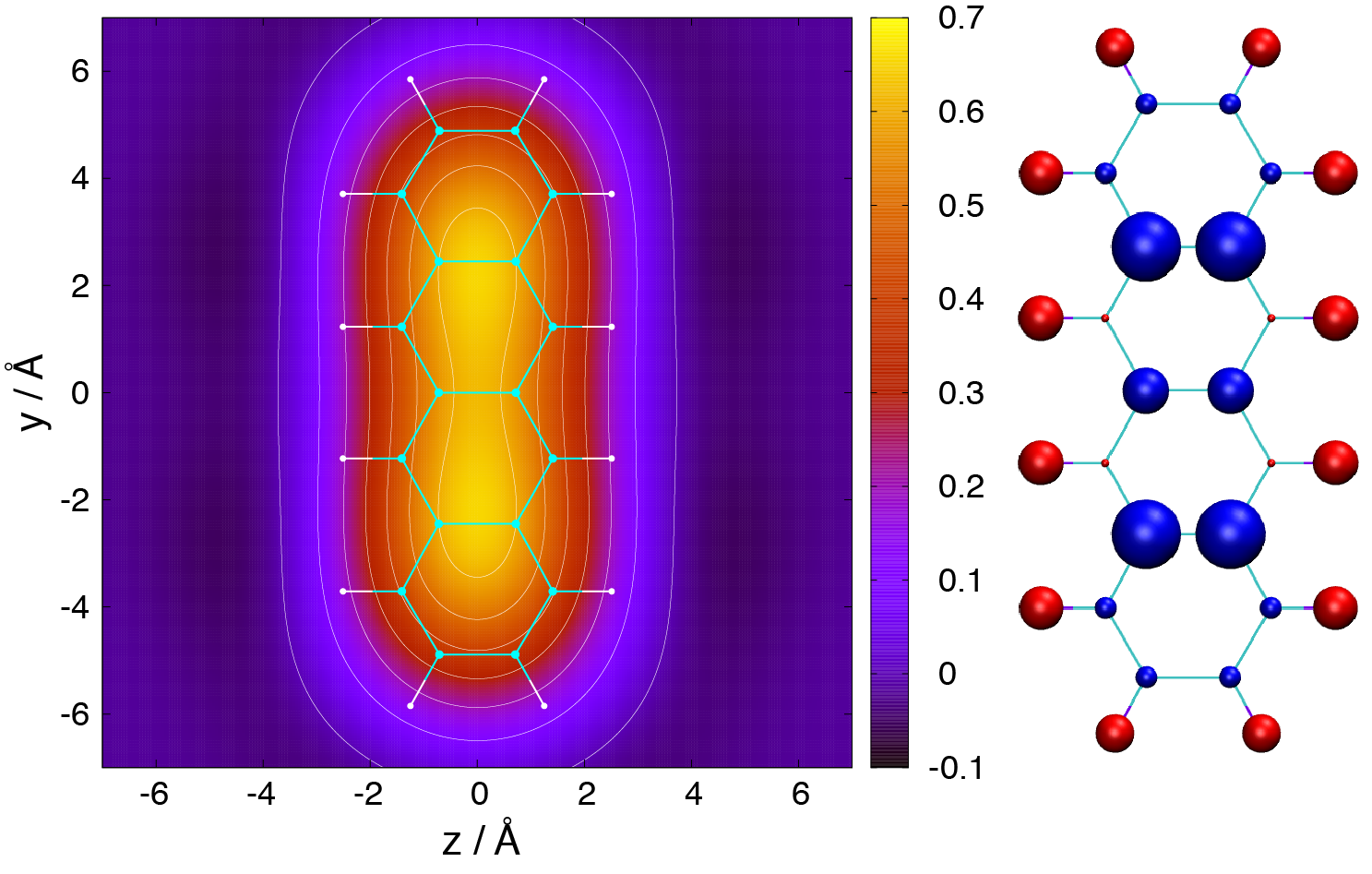}
    \caption{Left: The self-consistent polarization potential $V_{\rm pol}^{\rm SC}(r)$  (color bar units in eV, contour values: 0, 0.1, 0.2, 0.3, 0.4, 0.5, 0.6) for TET in parallel orientation and for   $R=100\ \rm{\AA}$ and       $d=1.0\ \rm{\AA}$.  Right: The change of Mulliken charges $\Delta q$; maximum at 0.0172~$e$.}
    \label{fig:TET_yz_pot}
  \end{center}
\end{figure*}

 The effect of the NP on the molecule's electron density is more pronounced if the latter faces the NP with its plane in the parallel orientation  as shown in figure~\ref{fig:TET_yz_pot}. Here, $V_{\rm pol}^{\rm SC}(r)$ is shaped by the negative partial charges at the C atoms (cf.~figure~\ref{fig:model_sys}a). The change of $V_{\rm pol}^{\rm SC}(r)$ across the molecule is about 0.5~eV, which causes a change in Mulliken charges of up to 0.02~$e$. Although small in absolute number, it corresponds to a relative change of about 40\%.

\begin{figure*}[tbh]
  \begin{center}
    \includegraphics[width=\textwidth]{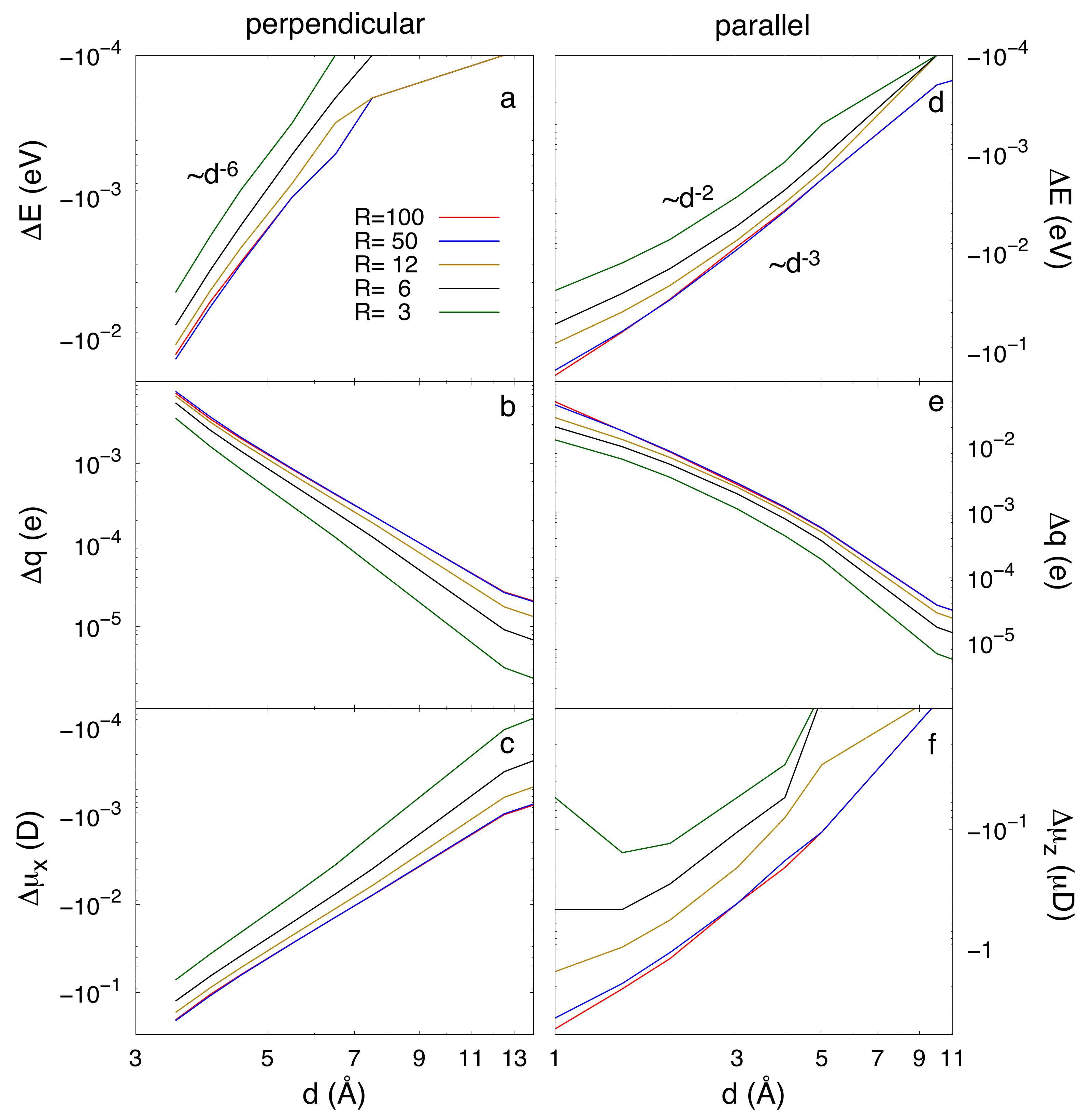}
    \caption{The dependence of interaction energy, $\Delta E$, change of Mulliken charge, $\Delta q$, and dipole moment, $\Delta {\bm \mu}$, on the distance, $d$, from the surface of the NP and its radius (in \AA) for the case of TET (note the double-logarithmic scale).}
    \label{fig:TET_delta}
  \end{center}
\end{figure*}

Figure~\ref{fig:TET_delta} shows the dependence of $\Delta E$, $\Delta q$, and $\Delta {\bm \mu}$ on the distance, $d$, from the surface of the NP for different NP radii. Overall, the influence of the NP on the molecular charge density is short-ranged and the effect on $\Delta E$ and $\Delta q$ is in magnitude larger for the parallel orientation as compared with the perpendicular one. The effect of the NP becomes essentially negligible once the distance to the closest atom exceeds about 5~\AA, independent on the radius of the NP. More specifically the distance dependence of $\Delta E$ is determined by the actual orientation. Fitting the distance dependence, one finds for short distances as shown in figure \ref{fig:TET_delta} $\Delta E \propto d^{-6}$ for $R=3$~\AA{} and 100~\AA{} in case of the perpendicular orientation. For the  parallel orientation it is $\Delta E \propto d^{-2}$  and $\propto d^{-3}$ for $R=3$~\AA{} and 100~\AA, respectively (see Supplementary Material (Suppl. Mat.), figure S1).

\textcolor{black}{Looking at the dependence on the NP radius, $R$, we notice that the interaction energy and the changes in the Mulliken charges decrease in magnitude with decreasing $R$. Closer inspection shows that in this case parts of the molecule are essentially too far away from the now much smaller NP, such that the respective charge densities do not contribute to the polarization field (for examples, see figure S2 in  Suppl. Mat.). We also notice that the effect of the NP on energy is converging with respect to the radius  once $R \gg L_{\rm mol}$, i.e. for the neutral TET, $R=100$~\AA{} already mimics the limit of an infinite plane.}

The quantum mechanical DFTB interaction energy, $\Delta E$, can be compared to the classical expression in equation~(\ref{eq:epol}). Differences occur for short distances only. For all perpendicular cases and around $d\approx 4-5$~\AA{} $E_{\rm pol}$ contributes above 95~\% to $\Delta E$, whereas for  the parallel cases it is above 90~\%.

The polarization field also causes a change in dipole moment as shown in panels c and f of figure~\ref{fig:TET_delta}. For the perpendicular orientation (plane of molecule is $xy$)  there is a change of the  $x$-component of the dipole due to the positive charge accumulation close the the NP (cf. figure~\ref{fig:TET_xy_pot}). For the parallel orientation (plane of molecule is $yz$) the dipole moment essentially does not change (note the scale in units of $10^{-3}$D), which is a consequence of the symmetry of the setup.
\begin{figure*}[tbh]
  \begin{center}
  \includegraphics[width=0.7\textwidth]{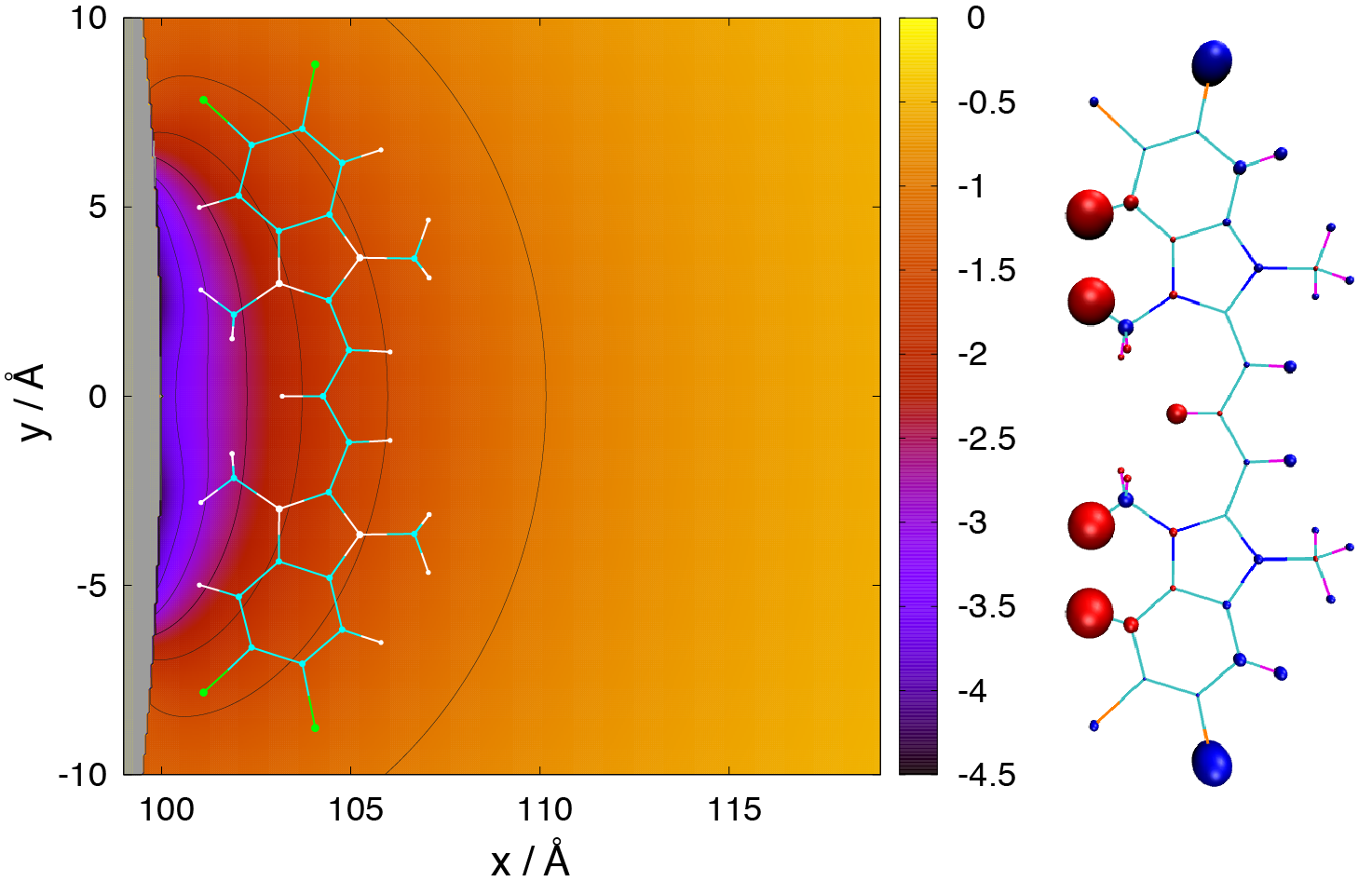}
    \caption{
    Left: The self-consistent polarization potential $V_{\rm pol}^{\rm SC}(r)$ (color bar units in eV, contour values: -3.5, -3, -2.5, -2, -1.5, -1) for CCY in perpendicular orientation and for a $R=100\ \rm{\AA}$ and       $d=4.0\ \rm{\AA}$.  Right: The change of Mulliken charges $\Delta q$; maximum at 0.0291~$e$.}
    \label{fig:ttbc100x1pot}
  \end{center}
\end{figure*}
\begin{figure*}[tbh]
  \begin{center}
    \includegraphics[width=0.7\textwidth]{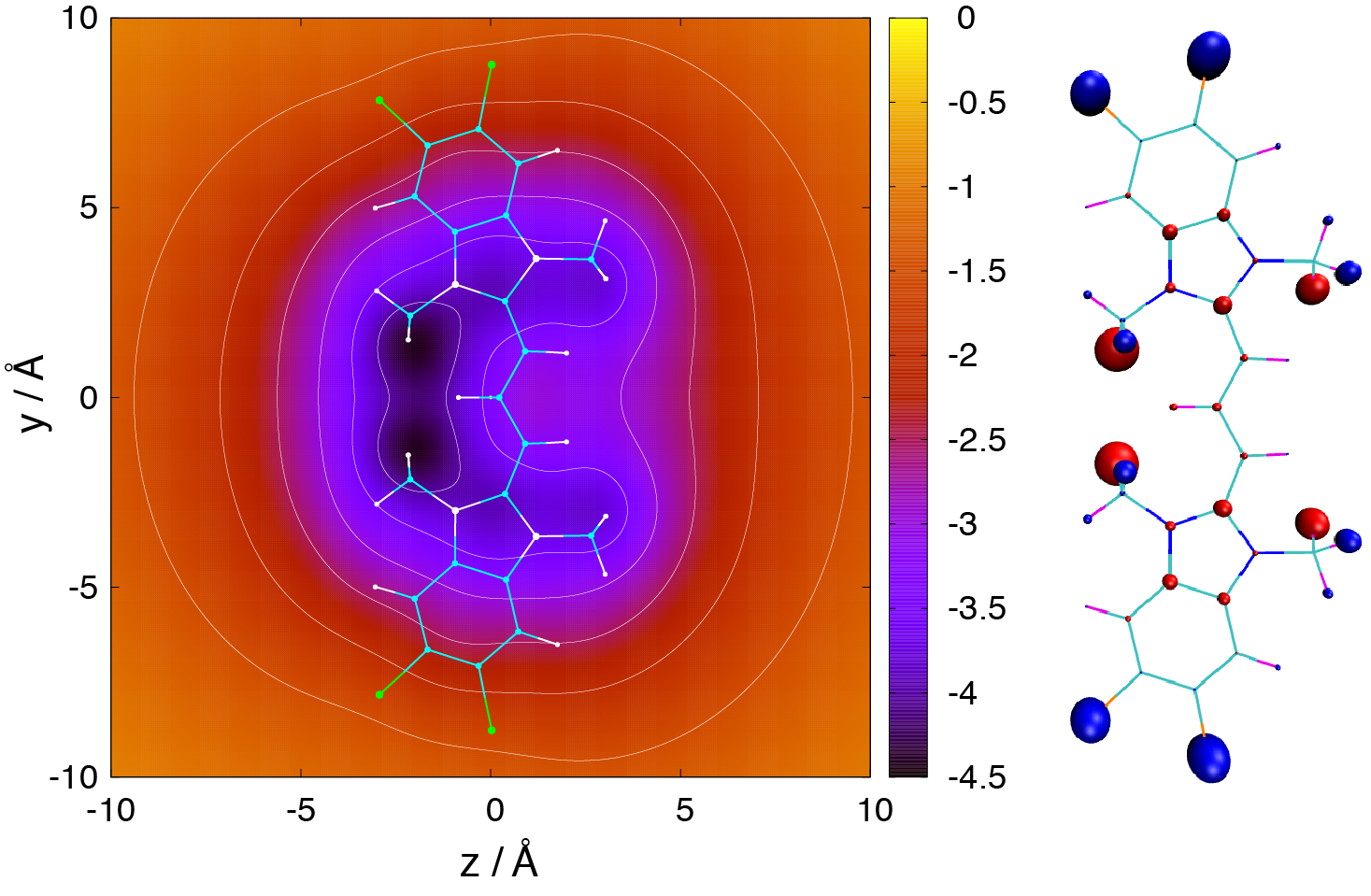}
     \caption{Left: The self-consistent polarization energy $E_{\rm pol}^{\rm SC}(r)$ (color bar units in eV, contour values: -4, -3.5, -3, -2.5, -2, -1.5) for CCY in parallel orientation and for a $R=100\ \rm{\AA}$ and       $d=1.9\ \rm{\AA}$.  Right: The change of Mulliken charges $\Delta q$; maximum at 0.0343~$e$.}
    \label{fig:ttbc100y1pot}
  \end{center}
\end{figure*}
\begin{figure*}[tbh]
  \begin{center}
    \includegraphics[width=\textwidth]{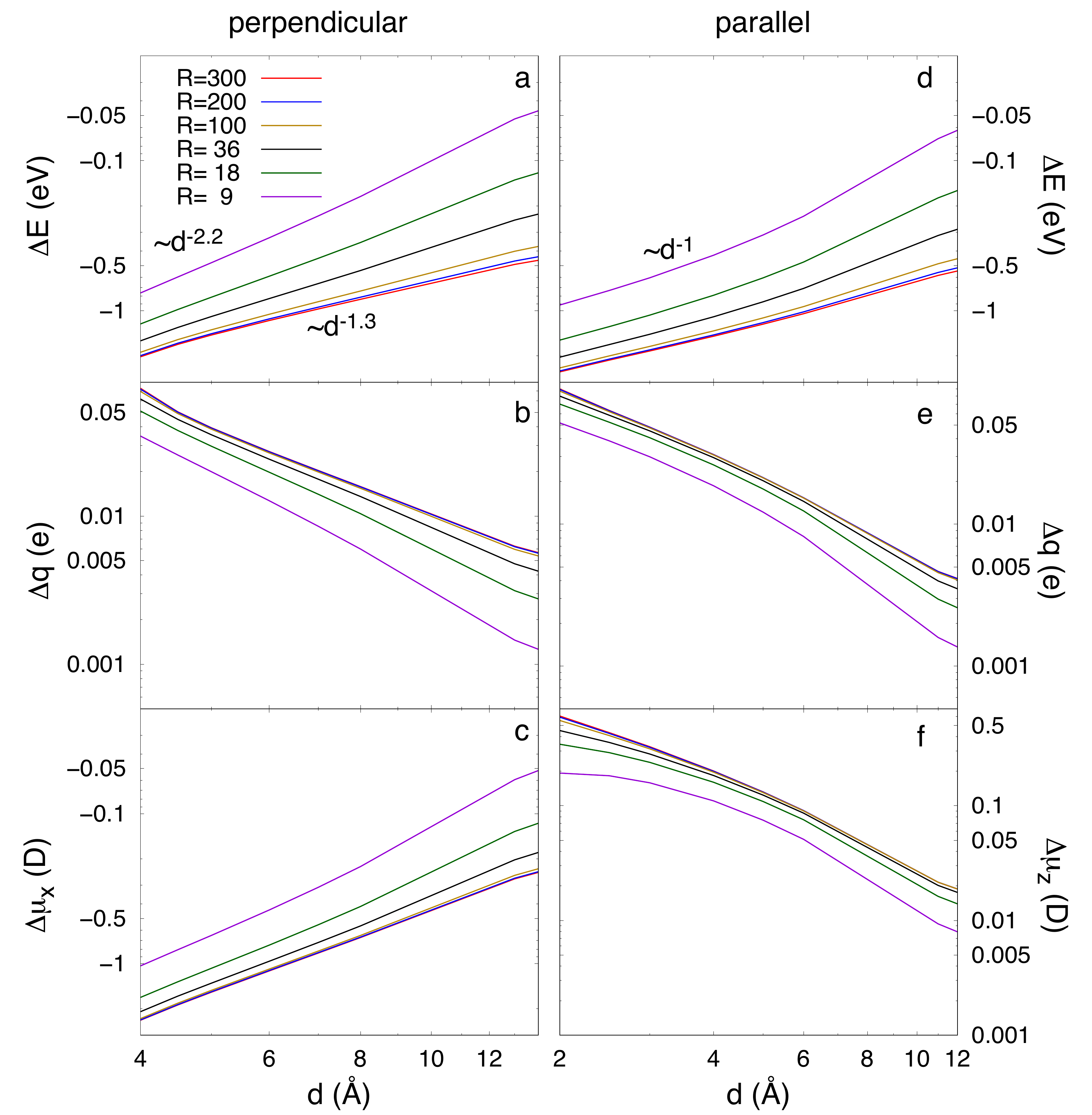}
    \caption{The dependence of interaction energy, $\Delta E$, change of Mulliken charge, $\Delta q$, and dipole moment, $\Delta {\bm \mu}$, on the distance, $d$, from the surface of the NP and its radius (in \AA) for the case of CCY (note the double-logarithmic scale).}

    \label{fig:ttbcdif}
  \end{center}
\end{figure*}
\subsection{Case of CCY}
The charge neutrality of TET rendered the effect of the NP on the electron density to be rather small. The situation is expected to change for CCY, which carries a net charge of +1~$e$.  \Fref{fig:ttbc100x1pot}. (left) shows the  polarization potential $V_{\rm pol}^{\rm SC}(r)$ for the parallel orientation  and for a NP with radius $R=100\ \rm{\AA}\gg L_{\rm mol}$  and $d=4\ \rm{\AA}$. Overall, the variation of the potential across the molecule is about 2~eV and the change in Mulliken charge is 0.029~$e$, i.e. 42~\%. The largest change occurs for the H-atoms, which are closest to the NP. Moreover, negative charge is accumulated at the Cl-atoms.

For the parallel orientation, where the results are shown in figure \ref{fig:ttbc100y1pot}, we first notice that in contrast to TET the polarization potential is of notable lower symmetry in accord with the different symmetry group the two molecules belong to. The potential changes by about 3~eV across the molecule and changes of the Mulliken charges reach 0.034~$e$ or 50\%. They are especially pronounced for those H-atoms which are out of the molecular plane and thus closest to the surface of the NP. In addition Cl-atoms acquire some extra negative charge.

Figure~\ref{fig:ttbcdif} shows the dependence of $\Delta E$, $\Delta q$, and $\Delta {\bm \mu}$ on the distance, $d$ from the NP for different  radii, $R$. First, we note that the effect of the NP is not as short-ranged as in the case of TET due to the presence of a net charge.  Further, for $\Delta E$ and $\Delta q$ the orientation is much less important than in the case of TET. Only the orientation dependence of $\Delta {\bm \mu}$ is as pronounced for CCY as it is for TET.

Inspecting the distance dependence of $\Delta E$ in the ranges shown in figure ~\ref{fig:ttbcdif} one finds by   fitting irrespective of the NP's radius  $\Delta E \propto d^{-1}$  in case of the parallel orientation. For the perpendicular orientation it is $\Delta E \propto d^{-2.2} $ and $\propto d^{-1.3}$ for $R=9$ \AA{} and 100~\AA, respectively (see Suppl. Mat., figure S1).

Concerning the dependence on the radius of the NP, the trend is rather similar to the case of TET, i.e. for NPs with $R \lesssim L_{\rm mol}$ parts of the molecular charge density  are located outside the range of effective interaction (for details, see Suppl. Mat., figure S2). 
\textcolor{black}{Convergence with respect to $R$ is reached for values exceeding about 200~\AA.}
For all perpendicular and perpendicular cases and around $d\approx 4-5$~\AA{} $E_{\rm pol}$ contributes above 95~\% to $\Delta E$. 
\section{Discussion of the Scaling Behavior}
\label{sec:scaling}
\textcolor{black}{
In the following we will focus on the behavior of the interaction energy as a function of distance from the NP, $\Delta E \propto d^{-n}$. According to the previous section this behavior is influenced by (i)  its orientation and (ii) the electronic properties of the molecule. First, we find a steeper distance dependence (i.e. larger values of $n$) in case of perpendicular orientation, irrespective of the molecule. Second, the exponent $n$ has systematically higher values in case of TET, indicating of different type of interaction.}

\textcolor{black}{
The different distance dependencies can be rationalized in terms of  a simple classical model. To this end the classical polarization energy, alike equation (\ref{eq:epol}), is considered for two cases, i.e. a single charge and a dipole oriented along the NP's surface normal. Results are calculated for varying distances of the corresponding charges from the center of the NP. The distance dependencies are then compared with the results obtained for an ideal uncharged metal sphere, which has the following analytical solution
\begin{equation}
  E_{\rm ideal} (r)= \frac{q q'}{4\pi \epsilon_0} \left( \frac{1}{r-r'}-\frac{1}{r}\right) \, .
\label{eq:ideal}
\end{equation}
Here, $q'=-qR/r$ is the image charge at position $r'=R^2/r$. Figure~S3 (Suppl. Mat.) shows that the results of the ideal model of  equation~(\ref{eq:ideal}) are recovered by the employed general description of the polarization field according to equation (\ref{eq:polarpot}) upon convergence with respect to the expansion order $l_{\rm max}$.}

\begin{figure*}[tbh]
  \begin{center}
    \includegraphics[width=0.99\textwidth]{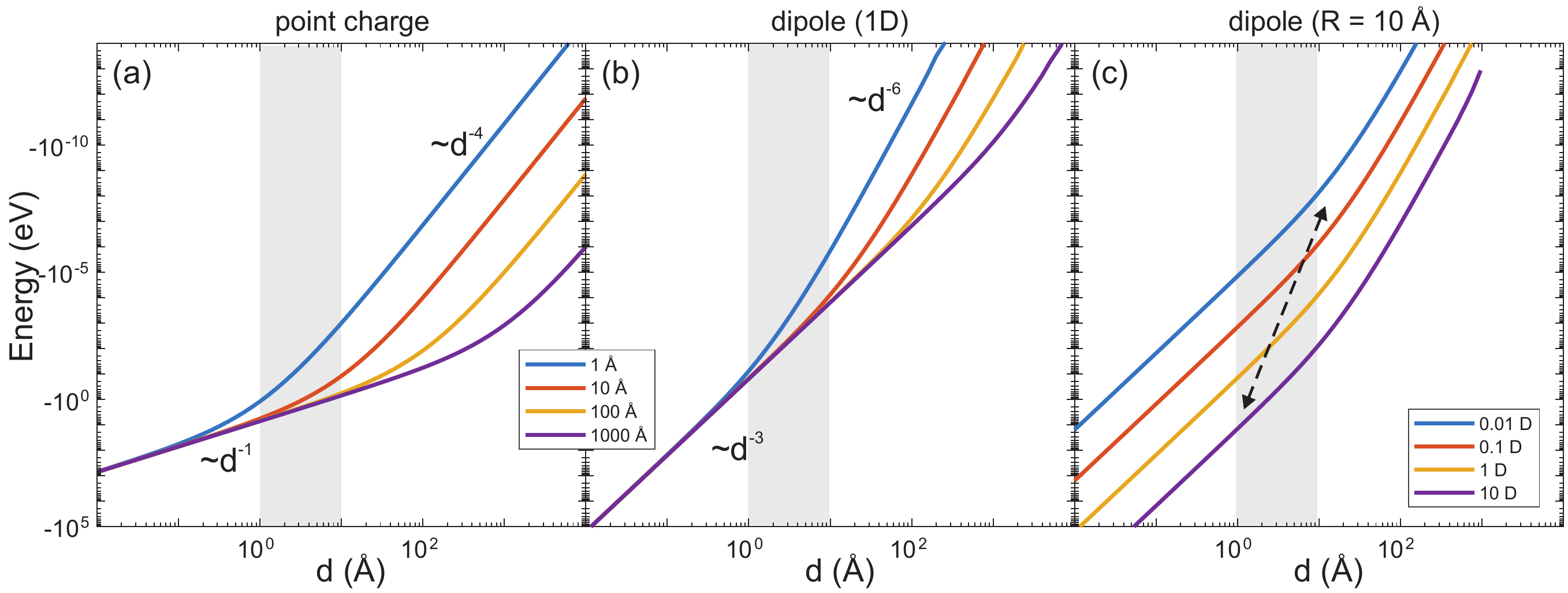}
    \caption{The dependence of the polarization energy,  equation (\ref{eq:ideal}), for a single charge $q=1~e$ (a) and a dipole of strength $1~D$ on the distance $d$ from the surface of the NP. The panels show results for different radii of the NP as indicated. In panel (c) the radius is fixed at $R=10$~\AA, but the magnitude of the dipole moment is varied as indicated. The shaded areas indicate the range relevant for the present TET and CCY model systems (Note the double-logarithmic scale.)}
    \label{fig:model_simple}
  \end{center}
\end{figure*}
\textcolor{black}{
Figure \ref{fig:model_simple} displays results for the two considered cases obtained using equation~(\ref{eq:ideal}). In both cases one notices a turnover behavior around $d\sim R$. For the case of a point charge (panel (a)) the distance dependence changes from  $\propto d^{-1}$ to $\propto d^{-4}$. 
Here, the former case corresponds to  a point charge interacting with its image inside the metal NP, whereas the latter case indicates the interaction of the charge with the dipole induced in the metal NP according to its polarizability.
In case of the dipole (panel (b)) one finds a transition from  $\propto d^{-3}$ to $\propto d^{-6}$ behavior.
For short distance this reflects the interaction between the actual and the image dipole. For long distances the actual dipole interacts with the induced dipole of the NP, in analogy to the case of a point charge. 
 }

\textcolor{black}{
Next we present a qualitative discussion of the comparison of the image charge model with the numerical results. We note that the distances covered by the DFTB analysis are ranging from  about 1 to 10 \AA, as indicated as grey-shaded areas in figure \ref{fig:model_simple}.
}

\textcolor{black}{
In case of CCY in parallel orientation we have roughly the $\propto d^{-1}$ behavior expected for charge-image charge interaction for $d<R$; cf. figure \ref{fig:model_simple}a. In perpendicular orientation, however, the value of the exponent $n$ increases  to about 1.2-2.2. This steeper distance dependence is attributed to an additional contribution from charge-induced dipole interaction. This conclusion is supported by the fact that a much stronger induced dipole is found in perpendicular orientation as compared to the parallel case, see figures~\ref{fig:ttbcdif}c,f.
}

\textcolor{black}{
 For TET in perpendicular orientation (figure~\ref{fig:TET_delta}a) the observed $\propto d^{-6}$ behavior seems to be in perfect accord with the expectation of the induced dipole interaction. Surprisingly, this behavior is found for the distance range $d<R$, in contrast to the prediction of $n=3$ from the simple model in  figure \ref{fig:model_simple}b. A possible explanation is indicated by  figure \ref{fig:model_simple}c,  showing that a change in  the magnitude of the dipole moment introduces an additional scaling of  the polarization energy. 
 In fact, the dipole moment in TET is of induced type with the actual magnitude being a function of the distance from the NP (cf. figure \ref{fig:TET_delta}c and figure S4 in the Suppl. Mat.). In other words, upon decreasing the distance $d$ the dipole moment increases, leading to a steepening of the distanced dependence, see the dashed  arrow in  figure \ref{fig:model_simple}c. 
}

\textcolor{black}{
In the parallel case there is essentially no induced net dipole moment (cf. figure \ref{fig:TET_delta}f) and no net charge, such that the simple model of equation~(\ref{eq:ideal}) would predict no effect at all. The fact that in figure~\ref{fig:TET_delta}d a distance dependence $\propto d^{-3}$ is found is attributed to the local structure of the electron density, see figure \ref{fig:TET_yz_pot}. At short distances the local charge distribution can be seen as a set of local dipoles, each giving rise to individual contribution with $\propto d^{-3}$ scaling.}

\section{Summary}
\label{sec:summary}
We have presented a proof-of-principle study, which combined a classical macroscopic description of the polarization potential of a NP with a quantum mechanical model for the charge density of a molecular system such as to quantify the effect of mutual polarization. Key to the model has been a discrete representation of the molecular charge density in terms of atom-based  Mulliken charges. They are obtained using DFT-based tight binding theory, where they constitute the basic quantity for the solution of the electronic Schr\"odinger equation. The latter has been linked to the polarization field in a self-consistent manner.

The approach was applied to two exemplary systems, containing dyes typically used for building supramolecular structures or molecular crystals. Whereas for the charge neutral TET the influence of the NP in the hybrid system has been moderate, the positively charged CCY showed rather pronounced effects. A quantitative comparison yielded a rather different dependence of the interaction energy on the distance from the NP's surface, but also on their mutual orientation.  In the perpendicular cases the interaction energy decays more rapidly for TET as compared with CCY. If the molecule's plane faces the NP (parallel case), the difference in distance dependence is only moderate.  The dependence on the radius of the NP is similar for the two dyes, i.e. the interaction energy decreases with decreasing radius, which is due to the fact that parts of the molecular charge density move effectively out of the range of interaction. It has also been noticed that the charged system shows a less pronounced orientation dependence of the interaction energy. The distance dependence has been contrasted with that of an ideal metal NP, interacting with a point charge and a dipole.  
\textcolor{black}{Upon inclusion of the peculiarities of the molecular charge density and its orientation and distance dependent changes, the qualitative  behavior is rather well explained by this simple model, i.e. in terms of interacting static charges/dipoles and induced dipoles.}

With DFTB being known as an efficient method to treat large systems, extensions of the present study to molecular aggregates or to include time-dependent external fields are within reach.
\section{Acknowledgments}
The authors thank the Deutsche Forschungsgemeinschaft (DFG) for financial support through the Sfb 652.

\providecommand{\newblock}{}

\end{document}